\documentclass[%
 reprint,
superscriptaddress,
 amsmath,amssymb,
 aps,
prb,
]{revtex4-2}

\usepackage{calrsfs}
\usepackage{graphicx}
\usepackage{dcolumn}
\usepackage{bm}
\usepackage{braket}
\usepackage[dvipsnames]{xcolor}
\usepackage{xcolor}
\usepackage{amsmath}

\def\beq{\begin{equation}}
\def\eeq{\end{equation}}
\def\bea{\begin{eqnarray}}
\def\eea{\end{eqnarray}}
\def\brcl{\begin{array}{rcl}}
\def\bccl{\begin{array}{ccl}}
\def\blcl{\begin{array}{lcl}}
\def\err{\end{array}}
\def\longeq{\;=\;}
\def\fatR{{\bf R}}
\def\fatF{{\bf F}}

\def\fatr{{\bf r}}

\begin{document}

\title{ 
Alchemical diastereomers from antisymmetric alchemical perturbations
}

\author{O. Anatole von Lilienfeld}
\email{anatole.vonlilienfeld@utoronto.ca}
\affiliation{Department of Chemistry, University of Toronto, St. George campus, Toronto, ON, Canada}
\affiliation{Department of Materials Science and Engineering, University of Toronto, St. George campus, Toronto, ON, Canada}
\affiliation{Vector Institute for Artificial Intelligence, Toronto, ON, M5S 1M1, Canada}
\affiliation{Machine Learning Group, Technische Universität Berlin and Berlin Institute for the Foundations of Learning and Data, Berlin, Germany}
\affiliation{Laboratory for AI and automation, Acceleration Consortium, University of Toronto. 80 St George St, Toronto, ON M5S 3H6}
\affiliation{Department of Physics, University of Toronto, St. George campus, Toronto, ON, Canada}
\author{Giorgio Domenichini}
\affiliation{University of Vienna, Faculty of Physics, Kolingasse 14-16, AT-1090 Vienna, Austria}
\affiliation{University of Vienna, Vienna Doctoral School in Physics, Boltzmanngasse 5, AT-1090 Vienna, Austria}
\date{\today}
\begin{abstract}
The energy difference between two iso-electronic systems 
can be  accurately approximated by the alchemical first order Hellmann-Feynmann derivative for the averaged Hamiltonian. 
This approximation is exact up to third order because even-order contributions cancel out.
This finding holds for any iso-electronic compound pair (dubbed `alchemical diastereomers'), 
regardless of differences in configuration, composition, or energy, 
and consequently, relative energy estimates for all possible iso-electronic alchemical diastereomer pairs, 
require only ${O}(1)$ self-consistent field cycles for any given averaging reference Hamiltonian. 
We discuss the relation to the Verlet algorithm, alchemical harmonic approximation (AHA) [\textit{J. Chem. Phys.}\textbf{162}, 044101 (2025)],
relative properties such as forces, ionization potential or electron affinities, and
Levy's formula for relative energies among iso-electronic systems 
that uses the averaged electron density of the two systems [\textit{J. Chem. Phys.} \textbf{70}, 1573 (1979)]. 
Numerical estimates accurately reflect trends in the charge-neutral iso-electronic diatomic molecule series with 
14 protons (N$_2$, CO, BF, BeNe, LiNa, HeMg, HAl), with systematically increasing errors. 
Using alchemical Hellmann-Feynman derivatives for toluene, we demonstrate the concept's broader applicability 
by estimating relative energies for all 36 possible alchemical diastereomer pairs from vertical iso-electronic 
charge-neutral antisymmetric BN doping of toluene's aromatic ring, with mean absolute errors of a few milli-Hartrees.
\end{abstract}


\maketitle

\section{Introduction}
Understanding the impact of stoichiometry and structure on properties is of fundamental concern, 
but it can also be beneficial for accelerating the discovery and design of materials and molecules.
The scope of generic screening attempts of chemical space, be it experimental or be it computational, 
is severely limited by the exponentially scaling wall that arises due to the combinatorial explosion of possible molecules or materials
that could be made for any given number of atoms.
The calculation of {\em absolute} solutions to the electronic Schr\"odinger equation, 
frequently obtained by numerically solving the variational problem for an approximated expectation value 
of a given electronic Hamiltonian, constitutes one of the major bottleneck when pursuing this goal.
Most use cases in chemistry and materials, however, involve only estimates of {\em relative} properties, 
eliminating the need to explicitly obtain absolute numbers.

\begin{figure}[!ht] \centering \includegraphics[width=\linewidth]{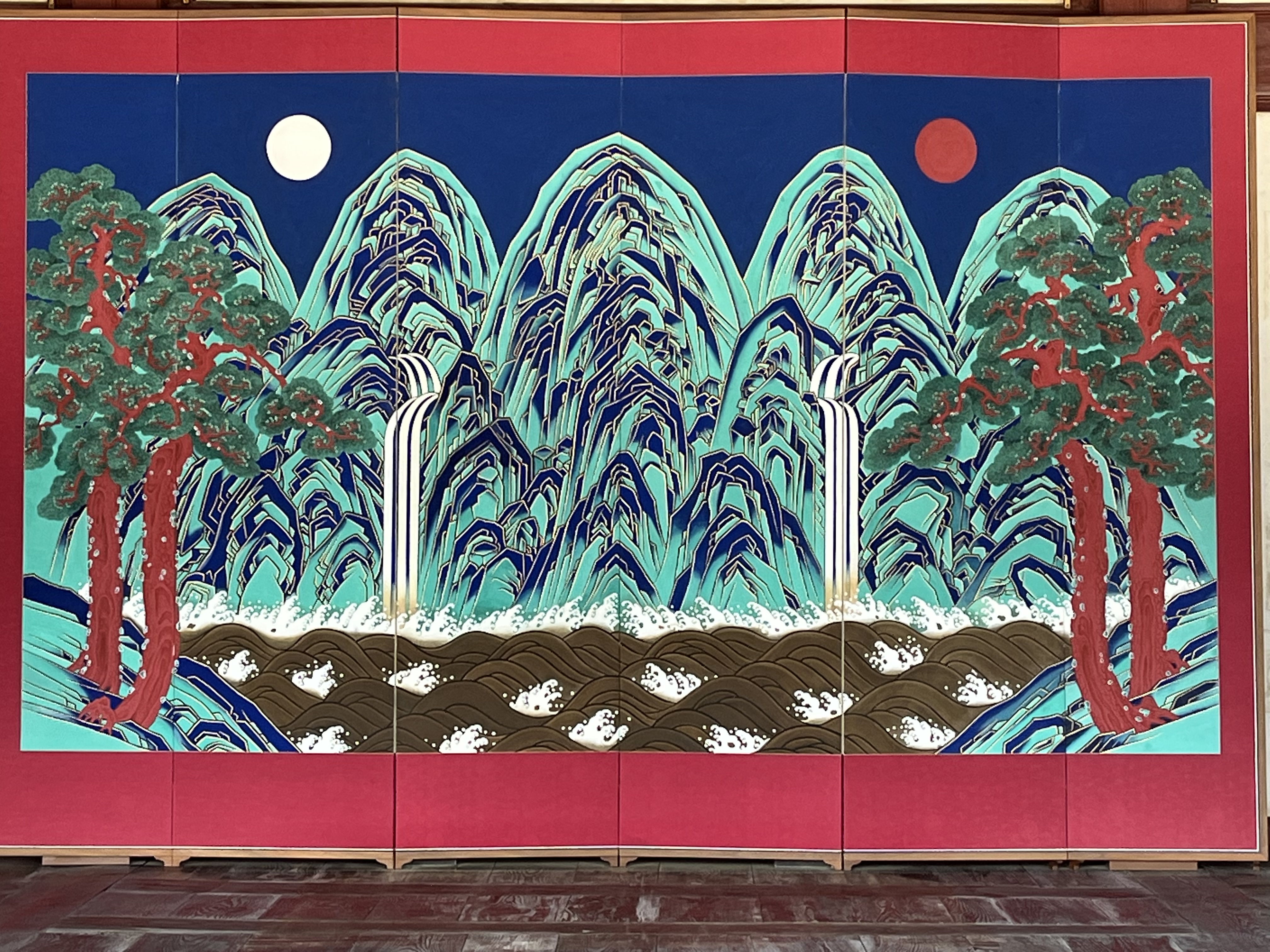}
    \caption{Historic Korean Irworobongdo painting 'Sun, moon, and five peaks' on folding screen 
    exhibiting imperfect symmetry.  
    Such illustrations created by unknown artists in the late Choseon era, 17th to 19th Century.
    'Sun, moon, and five peaks' was placed behind the king's seat in the Korean Joseon Dynasty.
    Photo picture taken by first author in 2022 in the main royal palace, Seoul, South Korea (built in 1395).
%
}
\label{fig:Fig0} \end{figure}

Alchemical perturbation density functional theory (APDFT)~\cite{APDFT} represents a computationally 
less demanding alternative approach to quantitatively estimate {\em relative} properties across chemical space.
APDFT, as any perturbative approach, is limited in accuracy by its radius of convergence within which 
sufficiently accurate estimates can be obtained, and which is dependent on the highest order considered.
Given modern implementations, APDFT estimates were shown to converge for predicted systems beyond commonly used
self-consistence field (SCF) convergence thresholds~\cite{von2021arbitrarily}.
APDFT relies on the continuous interpolation of (the nuclear charges in) the external potential. 
Expansions in the nuclear charge were already studied and exploited by the first adaptors of quantum mechanics in chemistry. 
Early relative computations, e.g. treating nuclear charges or entire functional groups as (non-discrete) parameters, 
trace back to Slater's rules, Hückel~\cite{hueckel_1931}, or Hylleraas \& Midtal~\cite{hylleraas_1956}.
Important contributions were made in the seventies and eighties, for example by 
Foldy~\cite{AtomicEnergyFoldy}, Wilson~\cite{WilsonsDFT}, Levy~\cite{levy1978energy,levy1979approximate},
Politzer and Parr~\cite{PolitzerParr}, or Mezey~\cite{ConcavityMezey1985}. 
Subsequent research in the nineties and 2000 dealing with alchemical changes includes, among others, 
Refs.~\cite{marzari_1994,AlchemicalDerivativeBinaryMetalCluster_WeigendSchrodtAhlrichs2004,anatole-prl2005,RCD_Yang2006,ArianaAlchemy2006,CatalystSheppard2010}, 
and more recently Refs.~\cite{LesiukHigherOrderAlchemy2012,CCSexploration_balawender2013,weigend2014extending,Samuel-JCP2016,StijnPNAS2017,munoz2017predictive,AlchemicalNormalModes,Keith2017alchemicalCatalysis,al2017exploring,balawender2018exploring,Samuel2018bandgaps,Keith2018benchmarkingalchemy,AtomicAPDFT,balawender2019exploring,shiraogawa2020theoretical,Keith2020acceleratedCatalystAlchemy,von2020rapidDeprotonation,munoz2020predicting,griego2021acceleration,gomez2021links,eikey2022evaluating,shiraogawa2022exploration,shiraogawa2022inverse,eikey2022quantum,shiraogawa2023optimization,balawender2023exploring}.

\begin{figure}[!ht] \centering \includegraphics[width=\linewidth]{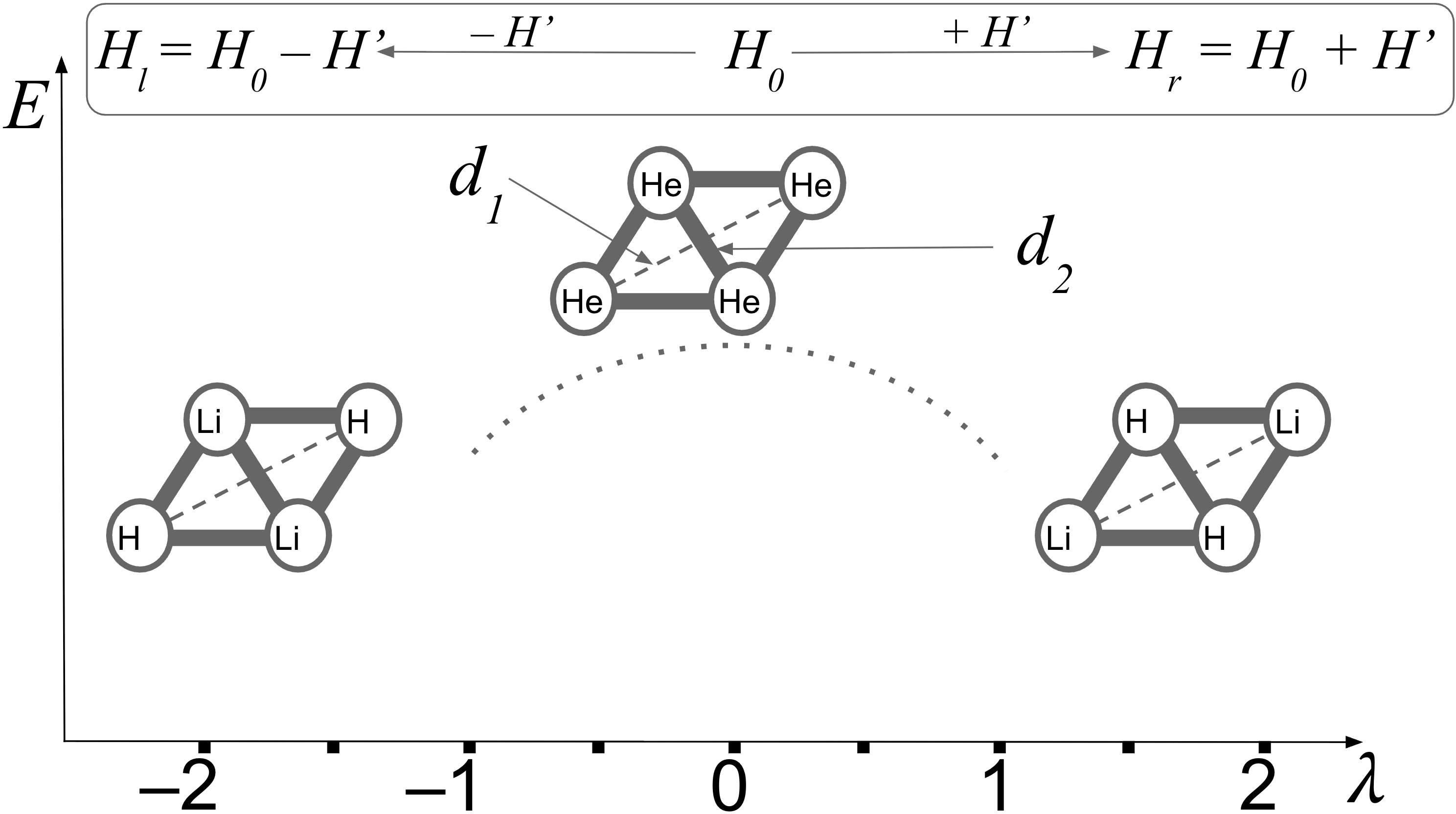}
    \caption{Qualitative drawing of the concave total potential energy (dotted) as a function of two 
anti-symmetric perturbations in the nuclear charges of an electronic reference Hamiltonian, $H_0$, in a planar coordinate
with two orthogonal interatomic distances $d_1$ and $d_2$.
Three exemplary chemical systems are shown for the fixed geometry of a planar rhombus 
with He$_4$ corresponding to the averaged reference Hamiltonian ($H_0$).  
For interatomic distances $d_1 = d_2$, the rhombus becomes a square for which
the chemical environments of all the atoms become identical in the reference system: 
the energy will be maximal for $\lambda = 0$, and (exactly) degenerate (due to rotational symmetry) 
for the two `alchemical enantiomers' at $\lambda = \pm 1$~\cite{AlchemicalChirality}.
For $d_1 \ne d_2$, the reference system's energy at $\lambda = 0$ will be shifted away from the maximum, 
lifting the degeneracy. The corresponding alchemical energy gradient at $\lambda = 0$ 
corresponds to the first order approximation of the energy difference between the two 
`alchemical diastereomers' at $\lambda = \pm 1$ [Eq.~\ref{eq:LD}].
}
\label{fig:FigDia} \end{figure}

\begin{figure}[!ht] \centering \includegraphics[width=\linewidth]{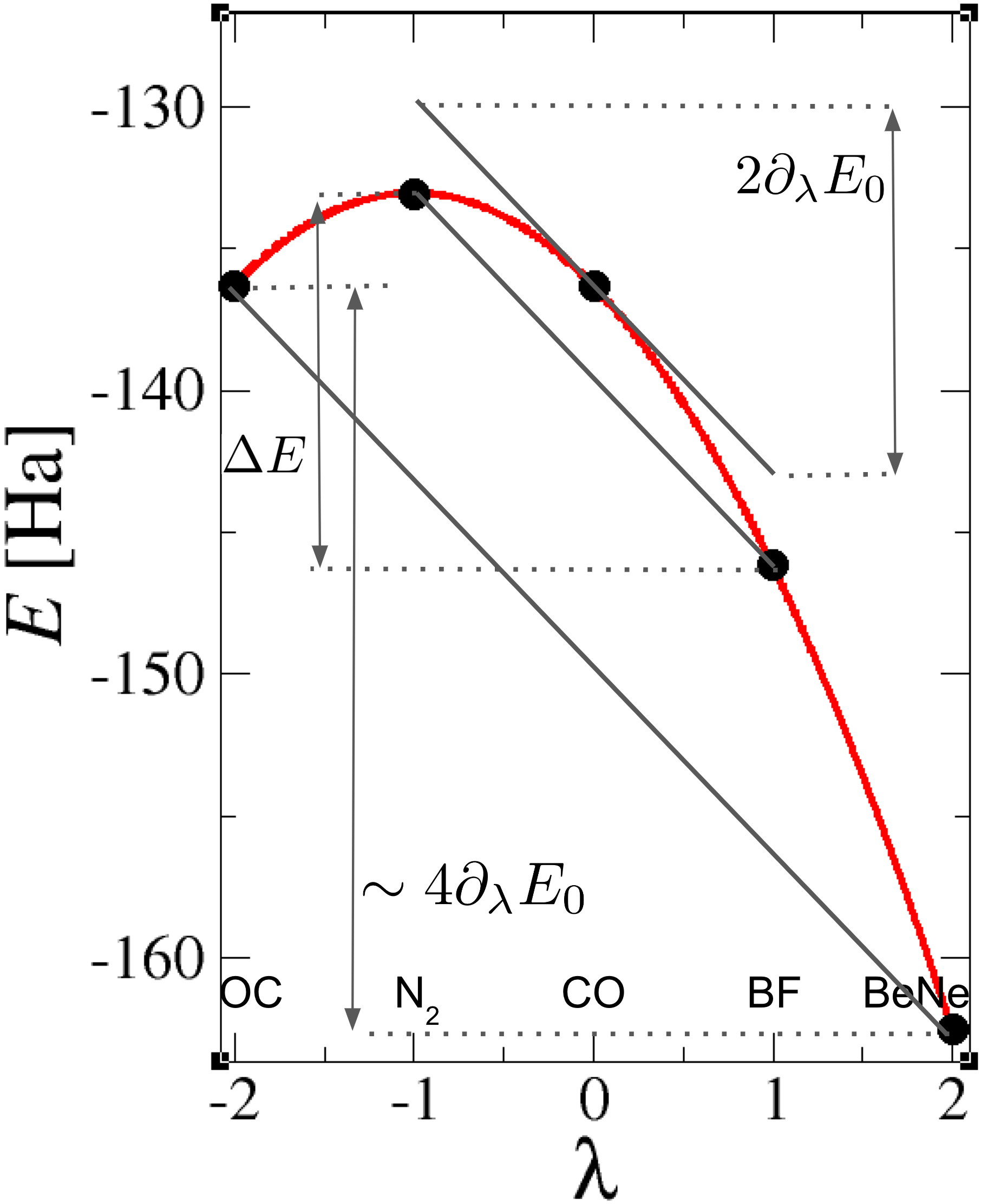}
    \caption{Electronic energy as a continuous function of linear coupling parameter $\lambda$ for some iso-electronic charge neutral diatomics with 14 protons.
N$_2$ and BF correspond to alchemical diastereomers of CO, implying that the leading order term of their energy difference $\Delta E$ corresponds
to twice the corresponding Hellmann-Feynman derivative evaluated for CO (See Eq.~\ref{eq:Relative}).
The relative energy between more distant alchemical diastereomers CO and BeNe is twice that amount. 
}
\label{fig:Fig1} \end{figure}

More qualitative considerations of alchemical changes can also be helpful. 
Alchemy has helped devising approximate covalent bond-strength models of similar performance 
as Pauling's electronegativity based model, yet with fewer parameters
and without the need for introducing additional properties which are not quantum observables, 
such as electronegativity~\cite{sahre2023quantum}.
More recently, the alchemical integral transform has led to 
simple approximate formulas relating solutions of one electronic Schr\"odinger equation  
to another~\cite{krug2022AIT,krug2024atoms,krug2025revAIT}.
While often crude and too coarse to reach chemically accurate predictive power, 
alchemy based models readily serve as baselines for $\Delta$-machine learning~\cite{DeltaPaper2015}, 
and have shown considerable gains in training data-efficiency~\cite{griego2020machine,krug2025AHA}.

The concept of alchemical chirality introduces approximate electronic energy degeneracies among seemingly
unrelated pairs of iso-electronic compounds (``alchemical enantiomers'')~\cite{AlchemicalChirality} that share the same geometry but differ in constitution or composition. 
For two compounds to correspond to alchemical enantiomers their external potentials have to average 
to an external reference potential with such symmetry that all atoms involved in transmutation have the same chemical environment.
In contrast to the regular spatial chirality and enantiomers, as commonly observed throughout nature for example in terms of the
handedness of amino acids, 
alchemical enantiomers have the same energy only up to 3rd order, i.e.~they are not {\em exactly} degenerate. 
As such, while the anti-symmetry in the alchemical perturbation Hamiltonian is exact, 
the corresponding eigenvalues are only approximately degenerate. 
In other words, alchemical chirality corresponds to a broken symmetry, as it is very common for symmetry elements in macroscopic 
samples, or even in the arts [see Fig.~\ref{fig:Fig0}].
Given the immense size of chemical space, we believe that also symmetry relationships that are merely approximate can still  
be useful, e.g.~to rank millions of compositional isomers as illustrated in Ref.~\cite{AlchemicalChirality}.

Here, we study the impact of lifting the requirement for the transmutating atoms to have the same chemical environment.
In particular, we consider anti-symmetric alchemical iso-electronic perturbations of {\em any} electronic reference Hamiltonian.  
The consequence of lifting the alchemical chirality symmetry requirement is that the first order term does no longer vanish. 
Since all even order contributions cancel out, however, alchemical chirality is recovered as the special case in which
the first order contribution disappears due to the similarity among atoms in the reference system involved in the anti-symmetric perturbation.
This idea is qualitatively illustrated in Fig.~\ref{fig:FigDia} for an example of a pair of alchemical diasteromers/enantiomers that
can be obtained when switching from a rhombus to a square in a planar He$_4$ reference system.


\section{Theory and methods}
\subsection{Vanishing even orders}
Consider two iso-electronic compounds $l$ and $r$ whose electronic Hamiltonian only differs in their respective external potentials, 
$\Delta v = v_r - v_l$.
Since corresponding changes of the non-relativistic concave electronic energy are conservative (energy being a state function), 
the force can be integrated in complete analogy to thermodynamic integration~\cite{TI}, 
$\Delta E = E_r - E_l = \int_{-1}^{+1} d\lambda \partial_\lambda E$
over a single one-dimensional coupling parameter, 
$-1 \le \lambda \le 1, \lambda \in \mathbb{R}$,
in the Hamiltonian, $\hat{H}(\lambda)$. 
Assuming a linear interpolation and defining the Hamiltonian such that the extreme values of $\lambda$ correspond to the
two respective compounds $l$ and $r$, 
$\hat{H}(\lambda) = (\hat{H}_r + \hat{H}_l)/2  + \lambda (\hat{H}_r - \hat{H}_l)/2$, 
we show below that the averaged mid-point Hamiltonian, $\hat{H}(\lambda = 0) = (\hat{H}_r + \hat{H}_l)/2 =: \hat{H}_0$,
represents an interesting reference system: 
It can serve as a pivot point for anti-symmetric perturbations in chemical compound space.

Consider the linear iso-electronic variation of the external potential, $\hat{H}' = (\hat{H}_r - \hat{H}_l)/2 = \delta v$.
One can think of $\hat{H}(\lambda) = \hat{H}_0 + \lambda \hat{H}'$ as a linear perturbation expansion 
around the average that is anti-symmetric along either direction of the path, i.e.~connecting to any two compounds $r$ and $l$
that happen to lie on the same $\lambda$ dimension in opposite directions and with $\hat{H}_0$ being exactly at their mid-point. 
Note how the corresponding two potentials, $v_l, v_r$ are respectively recovered
when setting $\lambda = -1, +1$, 
i.e.~$2\hat{H}' = 2\delta v = (v_r - v_0) + (v_0 - v_l)  = \Delta v$.

Also consider extending the $\lambda$ range further, e.g.
$-2 \le \lambda \le 2, \lambda \in \mathbb{R}$. 
In this case, another antisymmetric pair of iso-electronic (possibly fictitious) compounds is being coupled with the same
alchemically averaged Hamiltonian ($H_0$) at mid-point and corresponding to $\hat{H}_l$ and
$\hat{H}_r$ at $\lambda = -1$ and $\lambda = +1$, respectively. 
Generalizing this idea further, it is clear that multiple alchemical diastereomer pairs can be coupled using the same dimension in chemical space. 
For example, selecting the Hamiltonian of CO as the averaged reference system in the diatomic antisymmetric expansion series shown in Fig.~\ref{fig:Fig1},
pairs of alchemical diastereomers N$_2$/BF, CO/BeNe, BF/LiNa, BeNe/HeBe, LiNa/HAl, and HeBe/Si correspond to $\lambda$ = $\pm$1, $\pm$2, $\pm$3, $\pm$4, $\pm$5, and $\pm$6, respectively.

Assuming convergence, we can expand the respective electronic energy as a generic perturbation series using the averaged Hamiltonian
as reference, $E_0 = \langle \hat{H}_0 \rangle$, and involving antisymmetric variations towards positive or negative changes in $\lambda$.
More specifically:
\begin{widetext}
\begin{eqnarray}
E_r(\lambda)
& = & E_0 + \lambda \partial_\lambda E_0 + \frac{1}{2} \lambda^2 \partial^2_\lambda E_0  + \frac{1}{6} \lambda^3 \partial^3_\lambda E_0 
+ \frac{1}{24} \lambda^4 \partial^4_\lambda E_0 
+ \frac{1}{120} \lambda^5 \partial^5_\lambda E_0  
+ ... \\
E_l(-\lambda) & = & E_0 - \lambda \partial_\lambda E_0 + \frac{1}{2} \lambda^2 \partial^2_\lambda E_0  - \frac{1}{6} \lambda^3 \partial^3_\lambda E_0 
+ \frac{1}{24} \lambda^4 \partial^4_\lambda E_0 
- \frac{1}{120} \lambda^5 \partial^5_\lambda E_0 
+ .... 
\end{eqnarray}
\end{widetext}

Subtraction yields the energy difference between any two alchemical diastereomers, $\Delta E = E_r - E_l$, for which all even order contributions have vanished,

\begin{widetext}
\begin{eqnarray}
\Delta E (\lambda) &=&  2 \lambda \partial_\lambda E_0  + \frac{\lambda^3}{3} \partial^3_\lambda E_0  + \frac{\lambda^5 }{60} \partial^5_\lambda E_0 + ...  \nonumber\\
& = &  2 \lambda \int d\fatr \Delta v(\fatr) \rho_0(\fatr) + \frac{\lambda^3}{3}  \int d\fatr \Delta v(\fatr) \partial^2_\lambda \rho_0(\fatr)
+ \frac{\lambda^5}{60} \int d\fatr \Delta v(\fatr) \partial^4_\lambda \rho_0(\fatr) + ...
\label{eq:Relative}
\end{eqnarray}
\end{widetext}
where we are following Hellmann-Feynman's theorem~\cite{HF,WilsonsDFT,anatole-jcp2009-2,APDFT}, 
$\partial_\lambda E_0 = \int d\fatr \delta v(\fatr) \rho_0(\fatr)$, 
$\partial^3_\lambda E_0 = \int d\fatr \delta v(\fatr) \partial^2_\lambda \rho_0(\fatr)$,
$\partial^5_\lambda E_0 = \int d\fatr \delta v(\fatr) \partial^4_\lambda \rho_0(\fatr)$,
and where $\Delta v = 2 \delta v$. 
As already discussed in the context of alchemical perturbation density functional theory~\cite{APDFT}, 
these derivatives have clear meaning, summing up to the integral 
over the product of the perturbing Hamiltonian with the Taylor expansion in perturbed electron densities. 
Such expansions have been shown to rapidly converge for alchemical changes in fixed geometries and involving 
reasonably small variations in the nuclear charge distribution~\cite{von2021arbitrarily}.

\subsection{Alchemical enantiomers and diastereomers}
Note how this expansion recovers the case of alchemical chirality~\cite{AlchemicalChirality}
(where first order terms vanish) whenever the perturbing Hamiltonian and the 
reference system are chosen such that the parity of $\rho_0$ and anti-parity of $\Delta v$
result in a Hellmann-Feynman derivative (overlap integral
$\int d\fatr \rho_0 \Delta v$) that exactly averages out.
Given the reflection plane in the reference Hamiltonian's external potential,
as well as the energetic degeneracy up to third order, 
the iso-electronic compounds corresponding to $\hat{H}_l$ and $\hat{H}_r$ were dubbed `alchemical enantiomers'.
This alchemical chirality symmetry condition (resulting in a vanishing Hellmann-Feynman derivative) is not 
met, however, as soon as the transmutating atoms have chemical environments that differ. 
Consequently, non-vanishing values for the Hellmann-Feynman derivative at the reference 
system become the leading order contributions to energy differences of {\em arbitrary} iso-electronic compound pairs, 
and correspondingly, we dub the latter `alchemical diastereomers' for the remainder of this paper.
Conversely, alchemical diastereomers become enantiomers with approximate energy degeneracy 
in the limit of the relevant transmutating atoms possessing the same chemical environment in their averaged Hamiltonian. 

Fig.~\ref{fig:FigDia} exemplifies this point for the planar reference system He$_4$ for which 
the two planar constitutional isomers Li$_2$H$_2$ are diasteomers as long as the system remains a rhombus, i.e.~$d_1 \ne d_2$.
As $d_1 \rightarrow d_2$, the reference system's square symmetry is restored, and the energy of He$_4$ is at its maximum. 
Consequently, its alchemical Hellmann-Feynman derivative must be zero, 
and the two constitutional isomers Li$_2$H$_2$ correspond to alchemical enantiomers which, due to the rotational symmetry, are exactly degenerate.  
For example, if $d_1$ and $d_2$ respectively correspond to 3 and 2 {\AA}, calculating the electronic energy difference $\Delta E = E_r - E_l$ 
via the self-consistent field procedure (PBE0/cc-pVDZ) yields 0.731 Ha.
By contrast, 
the corresponding alchemical gradient based estimate [Eq.~\ref{eq:LD}] 
evaluated for the corresponding He$_4$ system ($\lambda = 0$) amounts to $0.708$ Ha, 
underestimating the SCF number merely by 23 mHa. 

Note that Eq.~\ref{eq:Relative} can easily be adapted to also estimate relative energies for arbitrarily distant 
alchemical diastereomers, simply
by increasing $|\Delta \lambda|$ to any other natural number as long as it is not larger than the smallest nuclear charge of a transmutating
atom in the reference Hamiltonian. 
Correspondingly, energy differences between alchemical diastereomers 
will grow linearly in $|\Delta \lambda|$ as long as they happen to be situated on the same $\lambda$ dimension in chemical space.
Fig.~\ref{fig:Fig1} illustrates this point: 
The energy difference between N$_2$ and BF is roughly half the size of the energy difference between CO and BeNe.

\subsection{Link to Verlet and the alchemical harmonic approximation}
Note how the exact antisymmetry of the perturbation leads to the exact cancellation of 
all the positive terms (even orders) in the left hand side expansion in Eq.~\ref{eq:Relative}, 
while the odd order terms are doubled.
Such alternations are also exploited in classical molecular mechanics, 
cf.~the time reversal symmetry within molecular dynamics simulation when using the Verlet algorithm: 
Velocity and higher odd order time derivatives cancel, and based on previous and current positions,
solely force based `classical' Newtonian propagation is exact up to 4th order~\cite{Moleculardynamics}. 
Accordingly, replacing the time variable $t$ by the alchemical coupling variable $\lambda$,  
and assuming that in addition to $E_0$ the alchemical equivalent to the previous position, $E_l$, was also known, 
one could estimate $E_r$ as follows
\bea
E_r & = & 2 E_0 - E_l + \lambda^2 \partial^2_\lambda E_0 + \frac{1}{12}\lambda^4 \partial^4_\lambda E_0 + ... 
\eea
where all odd order terms have vanished. 
Correspondingly, this approach might be beneficial when the first order perturbation of the electron density is available, 
for example in terms of the susceptibility kernel or from coupled perturbed self-consistent field calculation~\cite{LesiukHigherOrderAlchemy2012}),
which would yield $E_r$ exactly up to fourth order --- in exact analogy to the Verlet algorithm.

By contrast, the corresponding expression according to Eq.~\ref{eq:Relative} becomes
\bea
E_r & = & E_l +  2 \lambda \partial_\lambda E_0  + \frac{\lambda^3}{3} \partial^3_\lambda E_0  + ... 
\eea
with vanishing even order terms, 
odd derivatives being expanded at the averaged reference system, 
and with an off-set corresponding to energy of the left system. 
Equating and truncating both expressions yields, after rearranging, an expression for the curvature of $E$ in $\lambda$, 
\bea
\partial_\lambda^2 E_0 & \approx & 2 \frac{E_l - E_0}{\lambda^2} + \frac{2}{\lambda} \partial_\lambda E_0
\label{eq:AHA_curvature}
\eea
Note that this expression is consistent with the curvature identified within the alchemical harmonic approximation (AHA) 
which was published earlier this year (Eq.~13 in Ref.~\cite{krug2025AHA}). 
More specifically, Eq.~\ref{eq:AHA_curvature} generalizes the AHA curvature in the sense that the energy of the united atom is replaced by the energy of
any other system $H_l$ which happens to lie at $\lambda = \lambda_l$ on the same iso-electronic $\lambda$ coordinate in chemical space. 
This is considerably more convenient as it enables the freedom to specifically select $H_l$ such that the resulting AHA has maximal predictive power. 
The resulting generalized AHA expression then becomes,
\begin{widetext}
\bea
E^{\rm AHA}(\lambda) & = & 
(\lambda-\lambda_0)^2 \left[\frac{E_l - E_0}{(\lambda_l-\lambda_0)^2} + \frac{\partial_\lambda E_0}{\lambda_l-\lambda_0} \right]
+(\lambda-\lambda_0) \partial_\lambda E_0
+ E_0
\label{eq:AHA}
\eea
\end{widetext}

\subsection{Forces, Ionization potential, Electron Affinity}
The general usefulness of antisymmetric perturbations can also be shown for other energy differences, such as forces, 
ionization potentials or electron affinities.
Simple Taylor expansion for vertical alchemically induced  changes in forces on any atom in the system yields,
\bea 
\fatF_r  & = & \fatF_0 + \lambda \partial_\lambda \fatF_0  + \frac{1}{2}\lambda^2 \partial_\lambda^2 \fatF_0  + ... \\
\fatF_l  & = & \fatF_0 - \lambda \partial_\lambda \fatF_0  + \frac{1}{2}\lambda^2 \partial_\lambda^2 \fatF_0  + ... 
\eea
indicating that the first order change in $\fatF_0$ is of the same magnitude and in the opposite direction for the two diastereomers.
This might be a relevant constraint for improving alchemical geometry relaxations, 
e.g.~see Refs.~\cite{domenichini2022alchemical,shiraogawa2023optimization} for recently made contributions along such lines.
As a rule of thumb, one can see that the equilibrium geometry of the reference system can be expected 
to be approximately in between the relaxed geometries of the two alchemical diastereomers. 

Just as for the energy in Eq.~\ref{eq:LD}, the leading order for the difference in force change is simply twice the first order alchemical derivative of the force, 
which is given in terms of the perturbed electron densities for any atom $I$ exactly up to third order, 
$\fatF_{rI} - \fatF_{lI} \approx 2\lambda \partial_\lambda \fatF_{0I} + ... $. Here, the alchemical derivative of the force
is given by 
\bea
\partial_\lambda \fatF_{0I} & = & \int d\fatr \frac{Z_I(\lambda) \partial_\lambda \rho_0(\lambda) + \rho_0(\lambda) \partial_\lambda Z_I(\lambda)}{|\fatR_I-\fatr|^3} (\fatR_I-\fatr),
\nonumber \\
\eea
being directly related to the `alchemical force', i.e.~$\partial_{\fatR Z}^2 E$~\cite{AlchemicalNormalModes}.

Conversely, the ionization potential and electron affinity are defined as,
\bea
{\rm IP}  & = & E^+ -E \\
{\rm EA}  & = & E - E^- 
\eea 
where upper indices + and - denote removal and addition of one electron, respectively.  
Simple rearrangements and insertion of Eq.~\ref{eq:LD} yield, 
\bea 
{\rm IP}_r - {\rm IP}_l & = & 
E_r^+ -E_r - E^+_l+E_l \\
 & = & (E_r^+ - E^+_l)  -(E_r - E_l) \\ 
 & \approx & 2\lambda (\partial_\lambda E_0^+ - \partial_\lambda E_0), 
\eea 
and for EA correspondingly, EA $\approx  2\lambda (\partial_\lambda E_0 - \partial_\lambda E_0^-)$.
Note that these expressions could also be used to estimate trends in frontier molecular orbital (HOMO \& LUMO) eigenvalues, 
according to Koopman's theorem.
Preliminary numerical results are promising, 
suggesting that further in-depth studies might be warranted. 
While finalizing this manuscript, Shiraogawa et al.~have also derived similar relationships 
across chemical space for other response properties with promising predictive power~\cite{shiraogawa2025antisymmetry}.

\begin{table*}[!ht]
        \caption{
Relative electronic potential energy estimates in Ha between the diatomics to the left ($\hat{H}_l$) 
and to the right ($\hat{H}_r$) of the reference diatomic ($\hat{H}_0$). 
See Fig.~\ref{fig:Fig1} for an illustration.
$2 |\lambda| \partial_\lambda E_0 $ corresponds to the first order estimate (Eq.~\ref{eq:LD}), 
whereas Levy corresponds to Eq.~\ref{eq:Lev}, published by Levy in Ref.~\cite{levy1979approximate}.
All numbers obtained at fixed interatomic distance (1.1~{\AA}) and using the PBE0 functional in the
pcX-2 basis.
$\epsilon$ and $\tilde{\epsilon}$ correspond to the residual error of Eq.~\ref{eq:LD}, and its exponential fit.
} 
\begin{tabular}{l|l|r||r|r|r||r|r}
$\lambda_0 - \lambda_m$ & $\hat{H}_l$-$\hat{H}_0$-$\hat{H}_r$ &  $|\Delta E|$ &  $2 |\lambda | \partial_\lambda E_0 $ & $\epsilon$ & $\tilde{\epsilon}$  & Levy & $\Delta E -$ Levy \\\hline\hline
  & $|\Delta \lambda| = 1$ & & & & & \\\hline
1 & NN-CO-BF       & 13.176560  & 13.196927  & -0.020367 & -0.0163865 & 13.123976  & 0.052584 \\
2 & CO-BF-BeNe     & 26.207973  & 26.247951  & -0.039978 & -0.0390224 & 26.132007  & 0.075966 \\
3 & BF-BeNe-LiNa   & 39.024551  & 39.067086  & -0.042535 & -0.0648254 & 38.958482  & 0.066069 \\
4 & BeNe-LiNa-HeMg & 51.603454  & 51.669013  & -0.065559 & -0.0929274 & 51.482530  & 0.120924 \\
5 & LiNa-HeMg-HAl  & 63.762743  & 63.897974  & -0.135231 & -0.122873  & 63.490776  & 0.271967 \\
6 & HeMg-HAl-Si    & 75.074531  & 75.312538  & -0.238007 & -0.154374  & 74.768959  & 0.305572 \\\hline\hline
  & $|\Delta \lambda| = 2$ & & & & & &   \\\hline
2 & NN-BF-LiNa     & 52.201112  & 52.495903  & -0.294791 & -0.319473 & 51.669013  & 0.532099 \\
3 & CO-BeNe-HeMg   & 77.811426  & 78.134173  & -0.322747 & -0.53072  & 77.094902  & 0.716524 \\
4 & BF-LiNa-HAl    & 102.787295 & 103.338026 & -0.550731 & -0.760788 & 101.560490 & 1.2268 \\
5 & BeNe-HeMg-Si   & 126.677985 & 127.795948 & -1.11796  & -1.00595  & 124.707031 & 1.97095 \\\hline\hline
  & $|\Delta \lambda| = 3$ & & & & & & \\\hline
3 & NN-BeNe-HAl    & 115.963855 & 117.201259 & -1.2374   & & 112.968808 & 2.99505 \\
4 & CO-LiNa-Si     & 152.885958 & 155.007039 & -2.12108  & & 148.255308 & 4.63065 \\
\end{tabular}
\label{tab:Diatomics} 
\end{table*}

\subsection{Comparison to Levy}
Perturbing the electronic density for the averaged Hamiltonian,
one can also see the connection to Levy's estimation of iso-electronic energy differences 
\bea
\Delta E  &\approx  & \int d\fatr\, \Delta v(\fatr) \, \bar{\rho}(\fatr)
\label{eq:Lev}
\eea
that relies on the averaged electron densities, $\bar{\rho} = (\rho_r + \rho_l)/2$.~\cite{levy1978energy}
More specifically, assuming convergence for the left and right hand-side perturbation expansion,
\bea
\rho_r & = & \rho_0 + \partial_\lambda \rho_0  + \frac{1}{2} \partial^2_\lambda \rho_0  + \frac{1}{6} \partial^3_\lambda \rho_0 + 
\frac{1}{24} \partial^4_\lambda \rho_0
...  \nonumber \\
\rho_l & = & \rho_0 - \partial_\lambda \rho_0  + \frac{1}{2} \partial^2_\lambda \rho_0 - \frac{1}{6} \partial^3_\lambda \rho_0  + 
\frac{1}{24} \partial^4_\lambda \rho_0
... 
\eea
one can average and insert in Levy's formula which yields 
\bea
\int d\fatr\, \Delta v(\fatr) \, \bar{\rho}(\fatr)
& = &   \int d\fatr \Delta v(\fatr) \left(\rho_0  +  \frac{1}{2} \partial^2_\lambda \rho_0  + \frac{1}{24} \partial^4_\lambda \rho_0 + ...\right) \nonumber\\
\label{eq:Levy}
\eea
for which the even (odd) order energy (density) terms have also vanished.
Term-wise comparison to Eq.~\ref{eq:Relative} indicates that while Levy's formula correctly recovers
the first order energy term, it overestimates third, fifth, and seventh order energy terms respectively by factors 3, 5, and 7, etc.etc.
This implies that Levy's approximation should be less accurate than simply truncating after the first order term in Eq.~\ref{eq:Relative}. 
Applying Levy's formula for iso-electronic alchemical changes of atoms yields already a remarkable accuracy~\cite{levy1978energy}
which would indicate the dominance of the first order term.

\subsection{Computational Details}
Numerical calculations for the planar He$_4$ system (Fig.~\ref{fig:FigDia}) 
were done using Psi4~\cite{smith2020psi4} (version 1.9.1)
with the PBE0 density functional approximation~\cite{PBE0,PBE01,PBE02} 
and cc-pVDZ basis-set~\cite{ccpVDZ} (AUX). 

Numerical calculations for all the diatomics and toluene and its derivatives
were done using PySCF~\cite{pyscf_article,harris2020numpy,ipython} 
with the PBE0 density functional approximation~\cite{PBE0,PBE01,PBE02}. 
The importance of basis set effects for alchemical interpolations having 
been established previously~\cite{domenichini2020effects,domenichini2022alchemical},
we have used Jensen's pc2 basis set for hydrogens~\cite{pCn_jensen2001a,pCn_jensen2002a},
and the universal pcX-2 basis by Ambroise and Jensen for all other atoms~\cite{pcXbs}.
The interatomic distance of all systems in the charge neutral 14 electron diatomics series was set to 1.1{\AA}. 
The geometry of the 70 BN doped toluene derivatives was kept fixed to the
equilibrium geometry of toluene (given in Table \ref{tab:Toluene}),
obtained at the same level of theory and using Hermann's geometry optimizer PyBerny~\cite{PyBerny}.

In practice, the alchemical Hellmann-Feynman derivative can easily be calculated via chain-rule,
\bea
\partial_\lambda E & = & \sum_I \frac{\partial E}{\partial Z_I} \frac{\partial Z_I}{\partial \lambda}, \\
 & = & \sum_I \mu_I \Delta Z_I, 
\eea
where $I$ runs over all those atoms that are mutated.
Alchemical potentials, $\mu_I = \frac{\partial E}{\partial Z_I}$, have become readily available in many
quantum chemistry packages (e.g.~VASP~\cite{VASP}, Psi4~\cite{smith2020psi4}, MRCC~\cite{kallay2020mrcc}) 
in the form of total electrostatic potentials at the nuclear position $\fatR_I$--- 
in order to obtain the electronic contribution, one still has to subtract the nuclear repulsive contribution first. 
For this work, we have relied on the same python implementation as for the recently published work on relaxing molecular geometries via 
alchemical perturbations~\cite{domenichini2022alchemical} which is available on {\tt github}~\cite{Supplementary_QA}.

\begin{figure*}[!th] \centering \includegraphics[width=\linewidth]{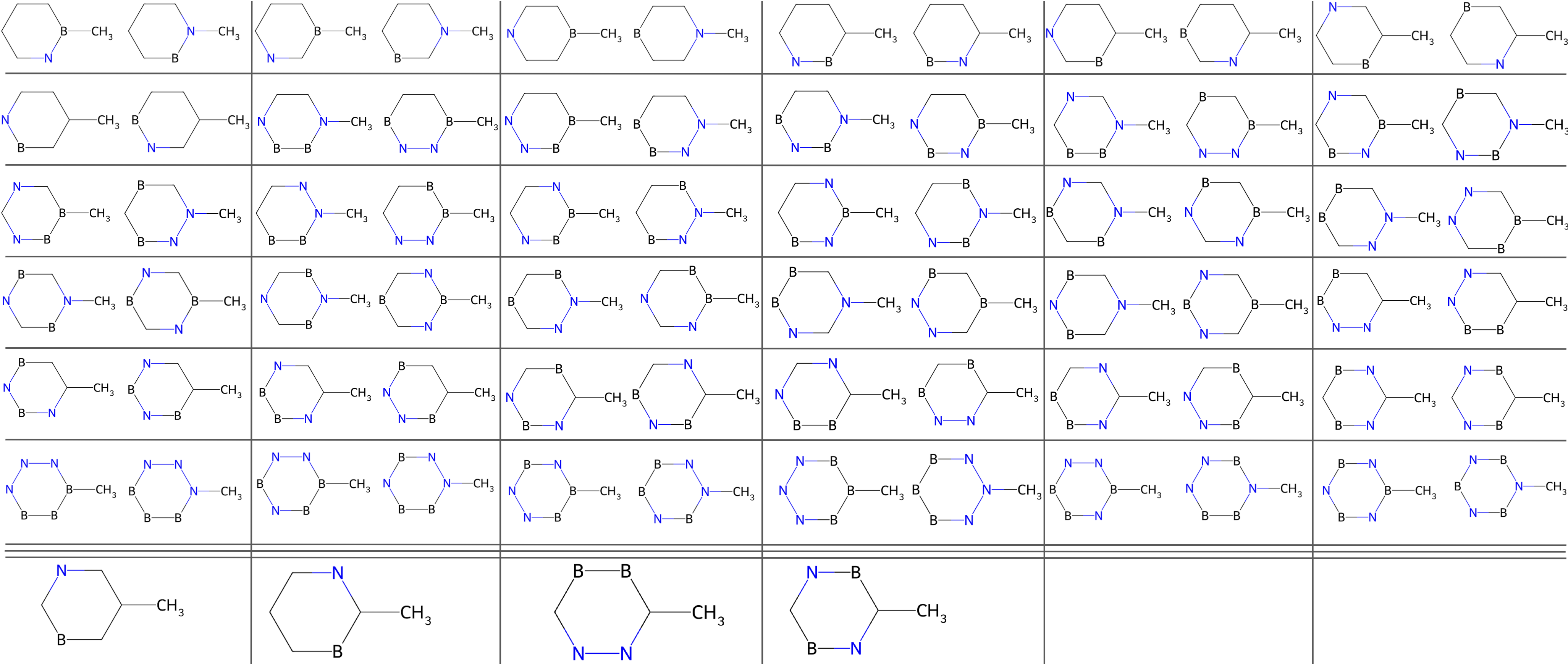}
    \caption{
Out of the 76 compositional and constitutional isomers one can obtain by iso-electronic charge-neutral doping of the aromatic unit in toluene with B and N,
 72 correspond to 36 pairs of alchemical diastereomers for which the leading order term of the
energy difference is twice the Hellmann-Feynman derivative evaluated for Toluene (Eq.~\ref{eq:LD}).
The diastereomer pairs are shown in the top 6 rows, the predictive power of Eq.~\ref{eq:LD} for the corresponding energy differences is shown 
in the scatter plot of Fig.~\ref{fig:Fig2}). 
The bottom line corresponds to the four toluene derivatives for which the corresponding anti-symmetry in the alchemical perturbations
aligns with the rotational $C_2$-symmetry axis in toluene, implying that no distinguishable anti-symmetric isomers exist. 
}
\label{fig:Tol} \end{figure*}
\begin{figure} \centering \includegraphics[width=\linewidth]{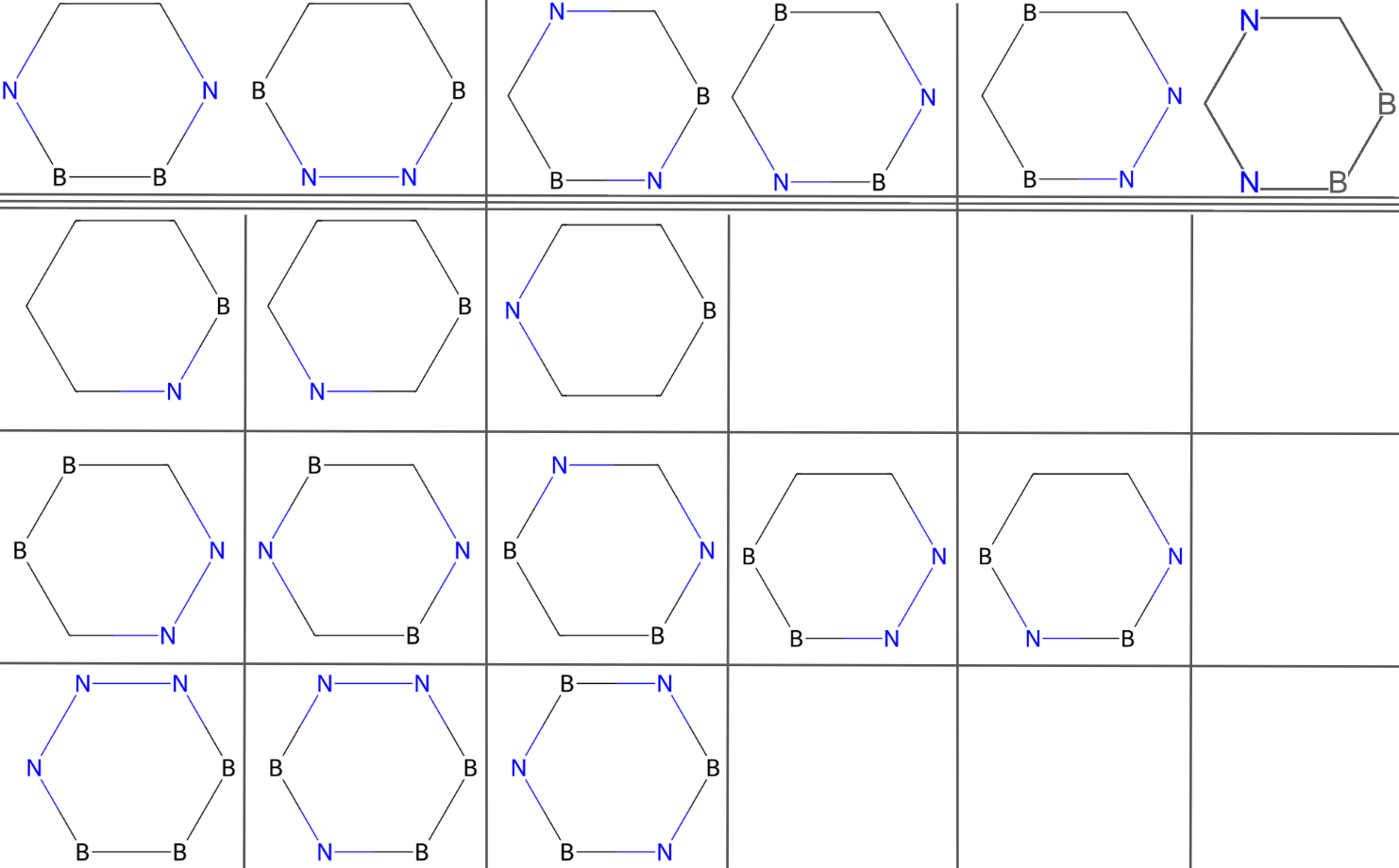}
    \caption{
All the 17 compositional and constitutional isomers one can obtain by iso-electronic charge-neutral doping of benzene with B and N.
The 6 derivatives in the top row correspond to 3 pairs of alchemical enantiomers (as introduced and discussed in Ref.~\cite{AlchemicalChirality}).
The remaining 11 derivatives (3 bottom rows) have at least one rotational symmetry element which eliminates any distinguishable anti-symmetric isomer. 
}
\label{fig:Ben} \end{figure}

\begin{figure*}[!th]
\centering \includegraphics[width=\linewidth]{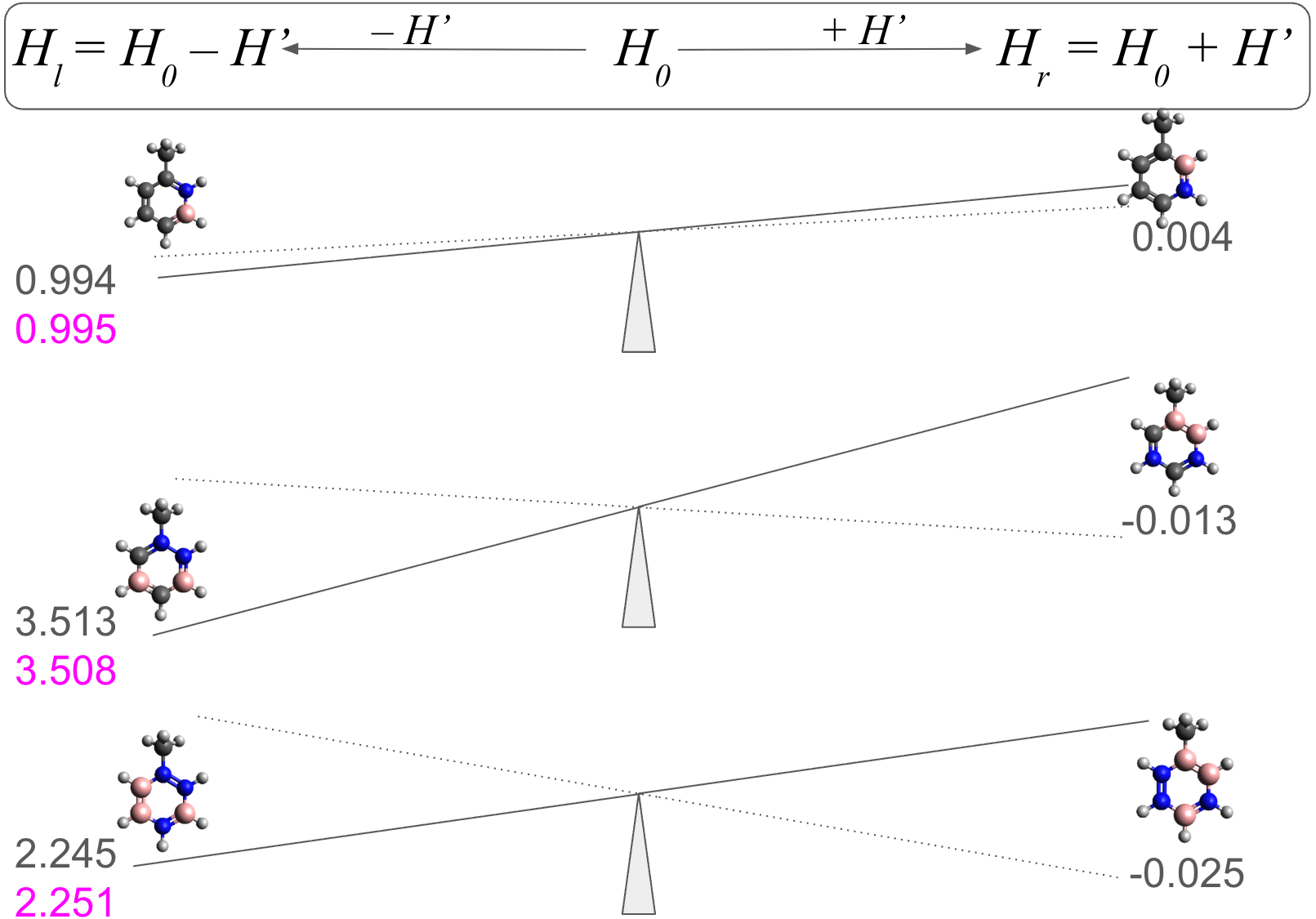}
    \caption{
Qualitative seesaw illustration of potential energy stabilization for 
three exemplary pairs of alchemical diasteomers obtained 
by antisymmetric BN doping of toluene (above the fulcrum). 
The three pairs were selected at random from all the 36 possible iso-electronic charge neutral coupling dimensions (Fig.~\ref{fig:Tol}) for which relative energy estimates are also shown in Fig.~\ref{fig:Fig2}. 
Numbers on the left and right respectively quantify the electronic (solid line) 
and total (dotted line) energy difference between the alchemical 
diastereomers drawn on the right ($E_r)$ and left ($E_l$), $\Delta E = E_r - E_l$ [Ha].
Numbers highlighted in pink correspond to the first order estimates of the 
electronic energy difference according to Eq.~\ref{eq:LD}. 
}
\label{fig:Fig3} \end{figure*}

\begin{figure}[!th] \centering \includegraphics[width=\linewidth]{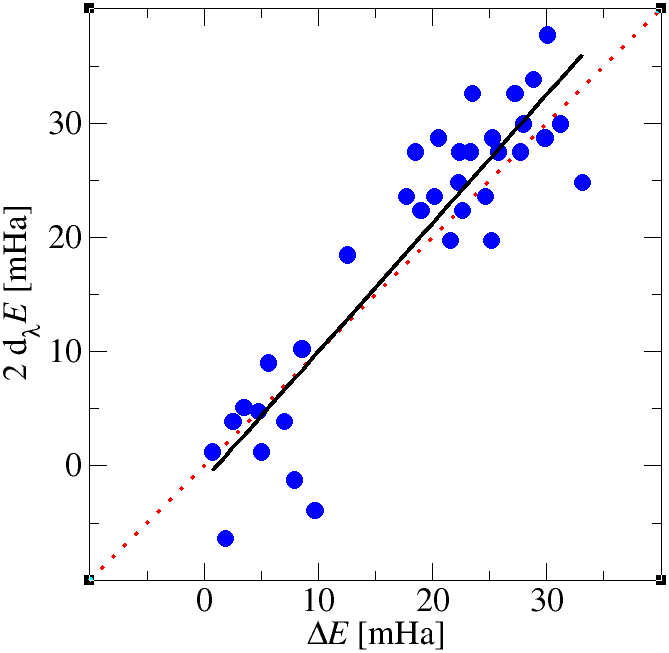}
    \caption{
Estimated total energy differences $2\partial_\lambda E_0$ (Eq.~\ref{eq:LD}) vs.~SCF based total energy difference ($\Delta E$) 
for all the 36 anti-symmetric BN doped alchemical diastereomer pairs of toluene as shown in Fig.~\ref{fig:Tol}.
On average Eq.~\ref{eq:LD} overestimates $\Delta E$ by nearly 1 mHa, and the mean absolute error (MAE) is 4.3 mHa. 
Linear regression (solid black line) further reduces the MAE to 3.3 mHa (correlation coefficient of 0.913).
}
\label{fig:Fig2} \end{figure}

\section{Numerical Results }
Truncating the expansion in Eq.~\ref{eq:Relative} after the leading order term, 
we have numerically evaluated the predictive power 
when estimating energy differences between alchemical diastereomers using 
\bea
\Delta E & \approx & 2 \lambda \partial_\lambda E_0 \longeq 2 \lambda \int d\fatr \Delta v(\fatr) \rho_0(\fatr)
\label{eq:LD}
\eea
Fig.~\ref{fig:Fig1} graphically illustrates this idea: 
Due to the antisymmetry condition of the perturbation, the energy's slope at the 
reference system provides a first order estimate of the energy difference
between the diastereomers. 
Note that this formula is consistent with the alchemical harmonic approximation (AHA) and that 
it becomes exact in the limit that the energy is parabolic in $\lambda$ (vide supra). 

\subsection{Neutral diatomics with 14 protons}
Following up on the discussion surrounding Fig.~\ref{fig:Fig1}, 
we have numerically evaluated Eq.~\ref{eq:LD}, as well as Levy's formula~\cite{levy1979approximate},
for all alchemical diastereomers in the charge-neutral iso-electronic 
diatomic series N$_2$, CO, BF, BeNe, LiNa, HeMg, and HAl, using step-sizes of 
$\lambda = \pm1, \pm2, \pm3$.
Results reported for a fixed interatomic distance of 1.1{\AA} in Table \ref{tab:Diatomics} 
indicate fair agreement, and deviate as little as 20 or 40 mHa when estimating the 
energy difference between N$_2$ and BF or CO and BeNe using the Hellmann-Feynman derivatives
for CO and BF, respectively.

As expected from above discussion, Eq.~\ref{eq:LD} provides a systematically 
better estimate of the actual energy difference than Levy's formula. 
This is encouraging since only one electron density is required for the averaged reference system, 
rather than two densities for each of the two end-points as it is necessary for Levy.
Furthermore, the same averaged reference system's electron density can be used to estimate energy 
differences between multiple pairs of compounds. 
As such, Levy's method scales linearly with number of compound pairs, 
while for our formula the scaling is constant. 

The residual deviation $\epsilon$ of $\Delta E$ from Eq.~\ref{eq:LD}, 
corresponding to all the higher order terms, is also shown in  Table \ref{tab:Diatomics}. 
Using only two fitting parameters, it can be approximated as
$\tilde{\epsilon} \approx a\,\exp((\lambda_0-\lambda_m)^b)$.
Linear least-square regression to the logarithmized form 
results in $\ln(a) = -4.3767$ and $b = 0.46801$ with a correlation coefficient of 
0.9809 and a mean absolute error of 87 mHa for the data given in Table \ref{tab:Diatomics}.
A mononomial fit in $\lambda_0-\lambda_m$, 
which would have been more in line with previous work~\cite{krug2024atoms,krug2025AHA},
yielded a slightly worse fit.
Estimates for the cases with $|\Delta \lambda | = 3$ were excluded from the fit due
to their extreme deviation well above 1 Ha. 
While the error is still substantial, the systematic nature of $\epsilon$ and
the good correlation of the fit suggest that it might be worthwhile to study the impact of 
the next third order term, or more sophisticated regressors.  

While results in  Table \ref{tab:Diatomics} correspond to a fixed bond-length of 1.1{\AA} only,
it is interesting to consider the impact on the predictive power when varying the distance. 
In particular, and as discussed in Ref.~\cite{krug2025AHA}, the alchemical curvature of the
electronic energy is expected to continuously interpolate between $\sim O(\lambda^{1/3})$ 
in the limit of $d \rightarrow \infty$ (assuming that the energy of the free
atom decays as $E(Z) \approx 0.5 Z^{7/3}$\cite{krug2024atoms}) 
and zero in the  $d \rightarrow 0$ limit (no $\lambda$ dependency for the united atom).

\subsection{BN doped derivatives of toluene}
We have selected toluene to further illustrate the potential of Eq.~\ref{eq:LD} for efficiently 
estimating relative energies in many dimensions. 
More specifically, we have evaluated Eq.~\ref{eq:LD}, 
for the 36 combinatorially possible $\lambda$-dimensions 
which cover all alchemical diastereomers that can be realized
through charge-neutral and iso-electronic BN substitutions of carbon pairs in the aromatic unit. 
All 36 pairs feature in Fig.~\ref{fig:Tol}, together with the 4 possible mutants for which
the anti-symmetry condition is trivially met because of the rotational $C_2$-symmetry axis
which is aligned with the carbon-carbon bond between the methyl group and the aromatic unit.
For comparison to a more symmetric reference molecule, we have also included the 
corresponding alchemical enantiomers in benzene in Fig.~\ref{fig:Ben}, as well as all the 
mutants that do not exhibit an alchemical symmetry-plane (see Refs.~\cite{AlchemicalNormalModes,AlchemicalChirality}).
Inspection of these two figures illustrates how a relatively modest reduction in 
the symmetry of the reference compound (insertion of a methyl group) leads to an
immediate combinatorial increase in the number of possible distinct alchemical mutants.
Notwithstanding the number of possible diastereomer pairs, however, 
Eq.~\ref{eq:LD} offers an effective way to estimate all the corresponding energy differences
without requiring any additional SCF cycles.

In the spirit of Fig.~\ref{fig:FigDia}, we have a drawn a seesaw diagram 
for three exemplary pairs of alchemical diastereomers of toluene in Fig.~\ref{fig:Fig3}.
Differences are shown for the true electronic energy, for the estimate in Eq.~\ref{eq:LD}, 
as well as for the total potential energy difference, i.e.~electronic and nuclear repulsion energy.
First, we note the reasonable agreement between the true electronic energy and the estimate based on Eq.~\ref{eq:LD}.
Furthermore, it is straightforward to interpret the trends in the sign: 
The diastereomer for which higher electron densities can be expected within closer
proximity typically exhibit a lower (more stable) electronic energy. 
This is the case for the examples shown whenever the nitrogen atoms (with more electrons)
are closer to each other, or closer to the electrons located at the methyl group. 
Note that this observation is also in line with previous findings about atomic energy contributions 
within Refs.~\cite{AtomicAPDFT,sahre2024transferability}. 
By contrast, addition of the nuclear repulsion typically reverses this trend due to lack of screening.

If sufficiently accurate, the usefulness of this approximation could be considerable: 
It would imply that given  the electron density of a reference systems, 
energy differences resulting from {\em any} of its combinatorially many possible
antisymmetric alchemical perturbations could be estimated with negligible overhead 
merely by evaluating the corresponding Hellmann-Feynman derivatives for each $\lambda$-dimension.
In other words, relative energy estimates for an arbitrarily large set of 
alchemical diastereomers, defined by their respective $\{\lambda_i\}$-perturbations,
can be generated for negligible additional computational 
cost---as long as they all share the same averaged reference Hamiltonian.
We have numerically exemplified this point for the BN doping of the aromatic moiety in the molecule toluene. 
In this case, BN doping defines 36 $\lambda$-dimensions space, 
along which we have estimated all the respective energy differences
between the corresponding alchemical diastereomers --- for `free', 
i.e.~via Eq.~\ref{eq:LD} and only based on the electron density obtained for the joint
averaged reference Hamiltonian of toluene.
Figs.~\ref{fig:Fig2} shows a scatter plot of the numerical estimates of energy
differences between 36 pairs of alchemical diastereomers of toluene.
The resulting MAE is 4.3 mHa ($\sim$2.7 kcal/mol), not far from chemical accuracy ($\sim$1 kcal/mol),
and similar to the accuracy of hybrid density functional approximations.

\section{Conclusion}
In conclusion, we have extended the concept of alchemical chirality to encompass alchemical antisymmetric perturbations, 
which effectively cancel even-order contributions to the relative energies of alchemical diastereomers, i.e.~iso-electronic
molecules which can differ not only in composition and structure but also in energy. 
The case of alchemical enantiomers—pairs of compounds differing in configuration and/or composition but exhibiting 
negligible energy differences—is naturally recovered when the averaged reference Hamiltonian possesses 
sufficient symmetry to nullify its alchemical Hellmann-Feynman derivatives. 
Our analysis also clarifies the interpretation of Levy's formula~\cite{levy1979approximate}, 
demonstrating that it corresponds to our first-order contribution, supplemented by all higher odd-order energy terms, 
which are systematically overestimated by a factor that scales linearly with order. 
Additionally, we have drawn parallels with the Verlet algorithm, established connections 
to the alchemical harmonic approximation~\cite{krug2025AHA}, and explored applications related to 
atomic forces, ionization potentials, and electron affinities.

Numerical evidence for energy differences among the 14 electron diatomic series, including N$_2$, CO, BF, BeNe, LiNa, HeMg and HAl,
indicates that the leading first-order term provides meaningful, and often accurate, estimates of energy 
differences between closely related alchemical diastereomers. 
The selection of the averaged reference system is critical, as it determines the dimensions of chemical space 
in which energy differences can be estimated with minimal computational cost. 
We illustrate this with 36 BN-doped alchemical diastereomers of toluene, where energy differences were 
computed using 36 Hellmann-Feynman derivatives based on a single electron density, achieving a relatively low mean absolute error of 4.3 mHa.

While finalizing this study, and going beyond the electronic ground-state energy, 
the concept of alchemical antisymmetry leading to compositional chirality and diastereomers has also been studied by 
Shiraogawa et al.~who derive simple relationships for response properties across chemical space~\cite{shiraogawa2025antisymmetry}.
All these findings are of immediate relevance to topics discussed and investigated in studies on large chemical spaces
spanned by poly-aromatic hydrocarbons, e.g.~by Chakraborty et al.~\cite{Raghu2019polycyclicBNdoped}, or in 
the COMPAS data-sets by Gershoni-Poranne and 
co-workers~\cite{wahab2022compas1,mayo2024compas2,wahab2024compas3,chakraborty2025compas4}.
Future studies could deal with excited states properties, 
the role of the quality of the electron density used within the Hellmann-Feynman derivative.
While the inclusion of higher order terms might also be considered for future research,
it is not obvious that the computational cost associated with evaluating the necessary second 
order electron density response~\cite{APDFT} for every alchemical dimension offers an
actual advantage over brute force screening. 


\begin{acknowledgments} 
The authors acknowledge discussions with M. Chaudhari, D. Khan, S. Krug, M. Meuwly, G. F. von Rudorff, M. Sahre, and A. Savin. 
Special thanks go to T. Shiraogawa, Y. Jung, and J. Schrier for substantial feedback.
We acknowledge the support of the Natural Sciences and Engineering Research Council of Canada (NSERC), [funding reference number RGPIN-2023-04853]. Cette recherche a été financée par le Conseil de recherches en sciences naturelles et en génie du Canada (CRSNG), [numéro de référence RGPIN-2023-04853].
This research was undertaken thanks in part to funding provided to the University of Toronto's Acceleration Consortium from the Canada First Research Excellence Fund,
grant number: CFREF-2022-00042.
O.A.v.L. has received support as the Ed Clark Chair of Advanced Materials, and as a Canada CIFAR AI Chair.
O.A.v.L. has received funding from the European Research Council (ERC) under the European Union’s Horizon 2020 research and innovation programme (grant agreement No. 772834).

Parts of this manuscript were paraphrased using GrokAI ({\tt http://www.grok.com}), 
however the authors are fully accountable for all statements made.

\end{acknowledgments} 
\bibliography{literatur} 

\appendix
\section{Toluene coordinates}
\begin{table}[!th]
    \centering
\begin{tabular}{lrrr}
Atom & X  & Y & Z \\
H &         -0.00048  &      2.76025 &       1.05495 \\
H &          0.88360  &      2.79792 &      -0.46619 \\
H &         -0.88315  &      2.79793 &      -0.46700 \\
C &         -0.00000  &      2.38982 &       0.02594 \\
C &          0.00001  &      0.89122 &      -0.00419 \\
C &          1.19407  &      0.17608 &      -0.00721 \\
C &         -1.19407  &      0.17608 &      -0.00705 \\
C &          1.19708  &     -1.21024 &      -0.00623 \\
C &         -0.00000  &     -1.90984 &      -0.00527 \\
C &         -1.19708  &     -1.21023 &      -0.00606 \\
H &          2.13561  &      0.71445 &      -0.01262 \\
H &          2.13883  &     -1.74590 &      -0.01129 \\ 
H &         -0.00000  &     -2.99283 &      -0.00868 \\
H &         -2.13883  &     -1.74589 &      -0.01100 \\
H &         -2.13560  &      0.71446 &      -0.01235 \\
\end{tabular}
    \caption{
Atomic coordinates in {\AA}~for toluene, as used to estimate energy differences between all alchemical diastereomers (See Fig.~\ref{fig:Fig2}). 
The geometry has been relaxed at the PBE0/pcX-2(Carbon)/pc2(Hydrogen) level of theory.
} \label{tab:Toluene} 
\end{table}
\end{document}